# Mechanical Properties of APbX$_3$ (A=Cs or CH$_3$NH$_3$; X=I or Br) Perovskite Single Crystals

Yevgeny Rakita, Sidney R. Cohen* [†], Nir Klein Kedem, Gary Hodes, David Cahen*

Materials and Interfaces Department and [†]Department of Chemical Research Support,
Weizmann Institute of Science, Rehovot. 76100, Israel

*corresponding authors: sidney.cohen@weizmann.ac.il ; david.cahen@weizmann.ac.il

## Abstract:

The remarkable optoelectronic, and especially photovoltaic performance of hybrid-organic-inorganic perovskite (HOIP) materials drives efforts to connect materials properties to this performance. From nano-indentation experiments on solution-grown single crystals we obtain elastic modulus and nano-hardness values of APbX$_3$ (A=Cs, CH$_3$NH$_3$; X=I, Br). The Young's moduli are ~14, 19.5 and 16 GPa, for CH$_3$NH$_3$PbI$_3$, CH$_3$NH$_3$PbBr$_3$ and CsPbBr$_3$, respectively, lending credence to theoretically calculated values. We discuss possible relevance of our results to suggested 'self-healing', ion diffusion and ease of manufacturing. Using our results, together with literature data on elastic moduli, we classified HOIPs amongst relevant materials groups, based on their elasto-mechanical properties.

Hybrid organic-inorganic perovskite (HOIP) - structured materials, mostly $CH_3NH_3PbI_3$ and $CH_3NH_3PbBr_3$, and the latter's inorganic analog $CsPbBr_3$[1] are of interest primarily for their remarkable photovoltaic properties[2]. The remarkable solar cell characteristics of HOIPs can be ascribed, at least in part, to the high charge carrier mobilities, especially for materials, prepared from solution at low temperatures. It has been speculated that this combination of features can be explained by "self-healing" or low kinetic formation barriers[3,4]. Such low barriers could support the suggested occurrence of ion migration, which has been invoked to explain hysteresis[5] observed in current–voltage characteristics of some of the devices [6–8]. The ability of HOIPs to support elastic local deformations is germane to these phenomena. However, despite the great interest in HOIPs in recent years, only limited information exists on their mechanical properties.

To try to clarify this matter, we carried out measurements and analyses of elasto-mechanical properties of single crystals of the methylammonium iodide and bromide and the all-inorganic Cs lead bromide; the corresponding iodide does not have a perovskite(-like) structure at STP. Because we include $CsPbBr_3$ in our study we will use the abbreviation *HaP*, *Ha*lide *P*erovskite, rather than HOIP, where appropriate. To investigate the mechanical properties, we performed nano-indentation on single crystals of $CH_3NH_3PbI_3$, $CH_3NH_3PbBr_3$ and $CsPbBr_3$ to determine their Young's Modulus (E) and nano-hardness (H), and combined those data with literature data of the thermal expansion coefficients (α) [9,10] of these and related materials. By comparing the data for HaPs to those for materials that are similar in structure, composition or functionality (photovoltaics) we can see if their mechanical behavior justifies associating them with known classes of materials. In general, classification of materials into groups provides insights into the common origin of observed physical properties and a basis to make predictions. In this light, classifications of the HaPs by their mechanical properties might give some idea about the nature of interatomic bonding and the flexibility and stability of the entire structure during manufacturing and operation of these materials. Finally, we try to shed light on the origin of these properties, of how they may affect their manufacturing and their relations to phenomena such as 'self-healing' and ion diffusion.

We use terminology such as stiffness (or the inverse term – compliance), which reflects the resistance to elastic deformation. Strong interatomic bonds cause higher resistance to elastic deformation, meaning – stiffer 'springs' that hold together the entire structure. *Young's, bulk and* *shear moduli* reflect the spring constants under uniaxial, volumetric and shearing deformations, respectively, and reflect the bond strength only in the elastic limit. The resistance of a material to plastic deformation can be determined from its *hardness*.

There are multiple ways to measure mechanical constants – compressing, pulling, bending, measuring the speed of sound, using an indenter for direct hardness measurements, measuring the lattice constants (diffraction) under pressure and others. The constants will reflect measured tension, compression, shearing or bending. Since we are interested in gaining understanding of inter-atomic characteristics via mechanical properties, the effects of long-range defects (e.g., dislocations, inclusions, cracks) need to be minimized. In addition we need data sets that allow extracting statistically meaningful data. Nano-indentation[11] allows measuring multiple data on single crystals with flat faces of sizes that can be readily grown.

## **Experimental:**

**Crystal Growth:** $CH_3NH_3PbI_3$ single crystals were grown following a published procedure[12], where a temperature gradient was used to grow the crystals on a silicon wafer that was partly immersed in a 57% hydroiodic acid solution of lead iodide and methylammonium iodide ($CH_3NH_3I$), where the part remaining outside the solution was cooled by an air flow. The resulting tetragonal (pseudo cubic)[13] crystals usually expose the (100) and (211) faces (see Figure 1).

For the growth of $CH_3NH_3PbBr_3$ we used anti-solvent (ethyl acetate) slow evaporation into a solution of N,N-dimethylformamide (DMF) of lead bromide ($PbBr_2$) and methylammonium bromide ($CH_3NH_3Br$). Single-crystal growth procedures of both $CH_3NH_3PbBr_3$ and of $CH_3NH_3Br$ have been described previously[4,14]. The resulting cubic $CH_3NH_3PbBr_3$[15] crystals always expose the (100) face (see Figure 2 (a)).

The growth method for $CsPbBr_3$ is similar to that used for $CH_3NH_3PbBr_3$ (see above), where the solvent is dimethyl sulfoxide (DMSO) and the anti-solvent is water or ethanol (to be published). The crystal structure was verified by comparing the powder-XRD to literature data[16]

All the precursor preparation and crystal growth were done under ambient conditions with 45%-65% relative humidity. None of the precursors or solvents were specially treated before use. All the crystals were stored in a desiccator in ambient with ~25% relative humidity.

**Nano-indentation:** Nanohardness and the indentation modulus were measured using an Agilent XP Nanoindenter. A Berkovich diamond indenter tip was loaded into the surface at a strain rate of 0.05 s$^{-1}$ to a depth of ~750 nm. The loading was done using the "Continuous Stiffness Measurement" (CSM$^{TM}$) mode [17] which gave the modulus and hardness continuously as a function of loading. The data were analyzed using standard Oliver and Pharr analysis[11] and averaged over 300-700 nm depths to avoid any surface effects. Two $CH_3NH_3PbBr_3$ and $CsPbBr_3$, and three $CH_3NH_3PbI_3$ samples were checked. 10-20 experimentally valid indentations were performed on each sample. The spacing between each indentation was 20 times the indentation depth (see Figure 2 (b)). A typical load vs. penetration depth curve is presented in Figure 2(c).

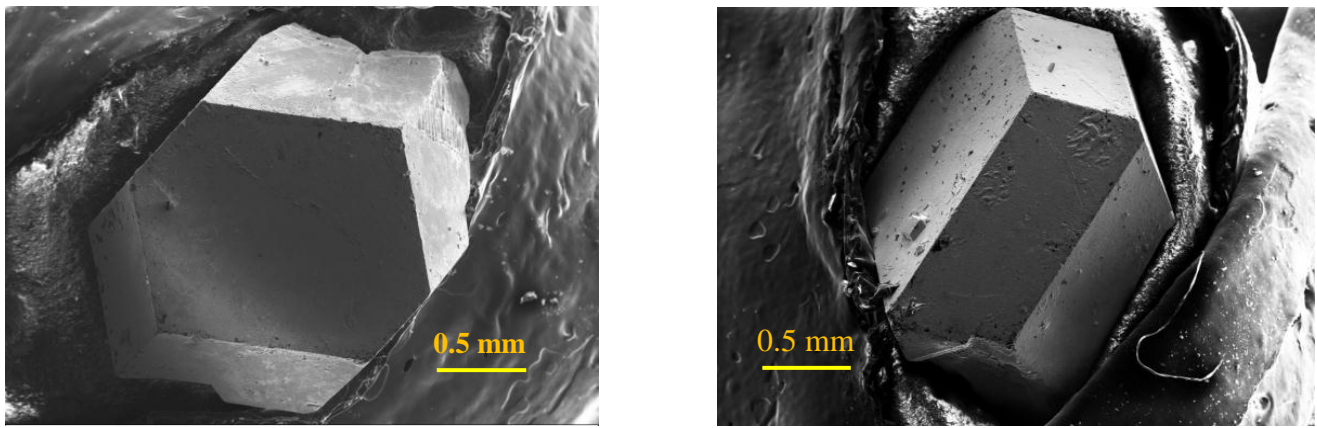

**Figure 1**: Scanning electron microscopy images of [left] (100) and [right] (112) faces of $CH_3NH_3PbI_3$ crystals.

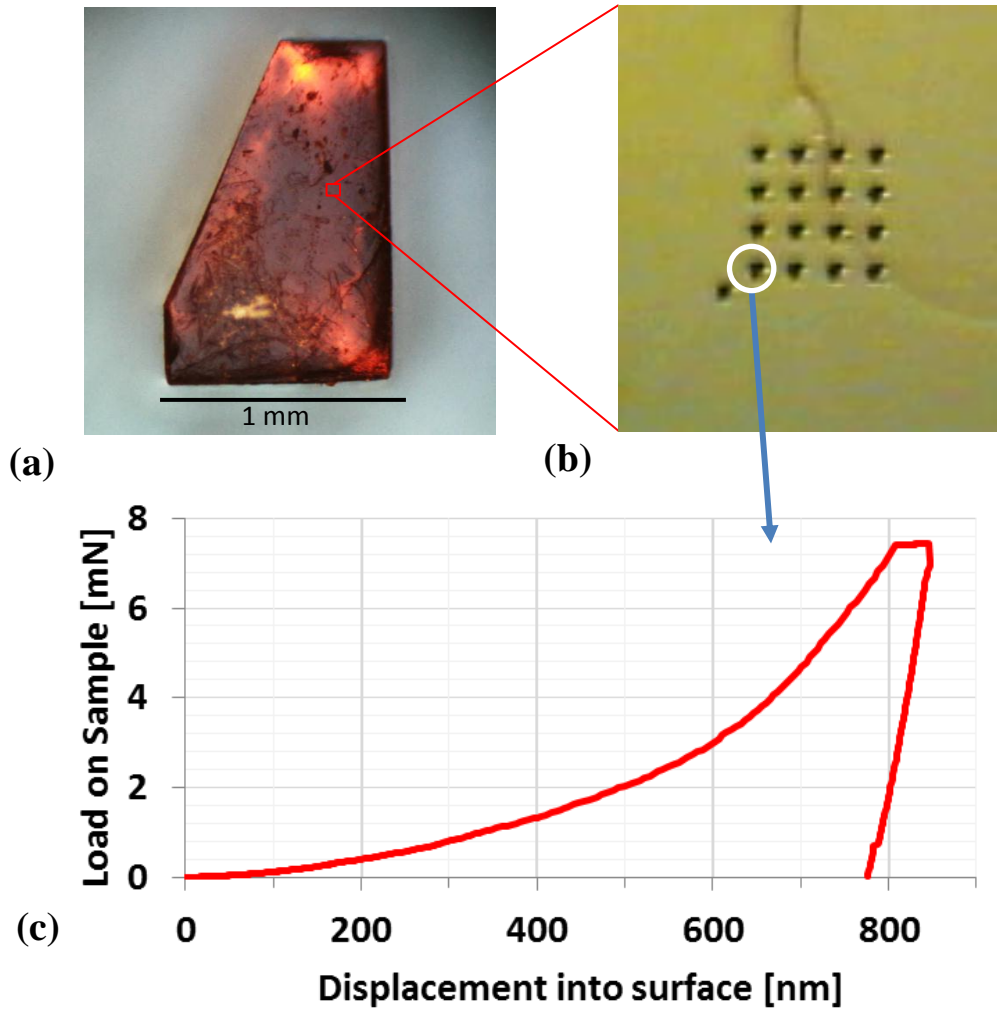

**Figure 2**: light-microscope images of (a) as-grown crystal of $CH_3NH_3PbBr_3$ presenting the (100) face and (b) image after indentation measurements. The spacing between the indentations is ~ 20 times the penetration depth of 750 nm.(c) Typical nano-indentation load/ displacement curve on $CH_3NH_3PbBr_3$.

**Results and Discussion:**

To the best of our knowledge, the data presented here comprise the first comparative study of H and E of $CH_3NH_3PbI_3$ and $CH_3NH_3PbBr_3$ single crystals. Table 1 lists these data, together with known physical properties. Bulk modulus, K, values for $CH_3NH_3PbBr_3$[18] and $CsPbBr_3$[19] were previously experimentally derived, and can be related to E through the Poisson ratio (see Table 1). The linear thermal expansion coefficients, α, of all three $APbX_3$ perovskite compounds were deduced from literature reports [9,18,20]. Where experimental information was missing we used results from theoretical calculations[21–23]. Materials in Table 1 were chosen for comparison, based on similarities in composition, crystal structure or functionality (i.e., photovoltaic materials). For comparison and as guideline for future measurements additional mechanical characteristics, bulk (K) and shear moduli (G), and Poisson ratio (ν) were derived using fundamental relations that strictly apply only to homogeneous, isotropic, and linear elastic materials[24].

Extensive theoretical work on the mechanical properties of several Pb and Sn -based HOIPs[22], which shows fair agreement with our experimental results, suggests that the type and strength of the Pb-X bond determine the elastic properties of the materials. Indeed, in agreement with the reported higher bond enthalpies for Pb-Br than for Pb-I,[25] we measure a higher Young's modulus for the Br-based HOIP than for the I-based one ($E_{(CH_3NH_3PbI_3)}$=14.2±1.9GPa and $E(CH_3NH_3PbBr_3)$=19.6±0.3GPa). Following this logic, the Young's modulus of 17 GPa [26] previously measured along a Pb-I network in $PbI_2$ -i.e., along the layers that make up the ($CdI_2$-type) structure, which lie in the c (0001) plane- is comparable to that of $APbI_3$ measured here. Even though $PbBr_2$ (average of 23.5 GPa [27]) has a STP structure that is very different from that of $PbI_2$ ($PbCl_2$-type; not layered), also here we find similar Young's moduli. This is consistent with the idea that the Pb-X framework dominates the mechanical properties of these materials (see Fig S3).

Comparison of the mechanical properties of the hybrid organic-inorganic with the fully inorganic $APbBr_3$ perovskite-structured material, reveals a slightly lower elastic modulus for $CsPbBr_3$ than $CH_3NH_3PbBr_3$ (the value for $CsPbBr_3$, derived from speed-of-sound measurements (2775 $\frac{m}{sec}$), was 26 GPa[19]). For $CH_3NH_3PbI_3$ and $CsPbI_3$ (for which only a value based on theoretical calculations is available), an opposite relation is seen, but, as $CsPbI_3$ does not have a perovskite(-like) structure at STP, such comparison may be invalid.

If we assume naively that the bond length reflects the bond stiffness (the shorter, the stiffer), then, based on the average Pb-X bond length of the $Cs^+$ and $CH_3NH_3^+$-based perovskites, the elastic modulus should have been higher for the hybrid than for the fully inorganic perovskite.

***Table 1***: *Mechanical constants of Halide Perovskites and related materials.*
The values that are measured in this work are in bold in lines 1-3. The remainder of the mechanical constants were derived [a,b] or taken from experimental or theoretical[23] results from other works, references to which are given in the Table. **E**- Young's Modulus; **H** – (nano) hardness; **G** – shear modulus; **K** – bulk modulus; **ν** - Poisson's ratio; **α** – linear thermal expansion coefficient; if only the volume value was available, it was divided by 3; **ρ** – density.

| Crystalline material | E [GPa] | H [GPa] | H/E | G [GPa] [b] | K [GPa] | ν | α [$10^{-6}$·$K^{-1}$] | ρ [gr/$cm^3$] |
|---|---|---|---|---|---|---|---|---|
| $CH_3NH_3PbI_3$ (100) face | **14.3(1.7)** | **0.57 (0.11)** | 0.040 | 5.4 | 13.9 [a] | $0.33^{22}$ | 43.3 [9] | $4.15^{36}$ |
| $CH_3NH_3PbI_3$ (112) face | **14.0 (2.0)** | **0.55 (0.12)** | | | | | | |
| $CH_3NH_3PbBr_3$ (100) face | **19.6 (0.3)** | **0.36 (0.01)** | 0.018 | 7.6 | 15.6 [18+(a)] | $0.29^{22}$ | 33.3 [10] | $3.80^{15}$ |
| $CsPbBr_3$ (101) face | **15.8 (0.6)** | **0.34 (0.02)** | 0.022 | 5.9 | 15.5 [(a)+(e)] | 0.33 [c] | 37.7 [20] | $4.83^{16}$ |
| $CH_3NH_3SnI_3$ | $16.6^{(a)}$ | --- | --- | 6.5 | $12.6^{37}$ | $0.28^{22}$ | --- | $3.65^{37}$ |
| $CsPbI_3$ (non-perovskite) | $20.1^{(a)}$ | --- | --- | 7.9 | $19.8^{(a)}$ | 0.33 [c] | --- | $5.39^{36}$ |
| $PbI_2$ - c (0001) plane | 17 [26] | $0.82^{26}$ | 0.048 | --- | --- | --- | $40^{38}$ | 6.16 |
| $PbBr_2$ [d] | 23.5 | --- | --- | 8.8 | 23.7 [a] | 0.33 | $31.7^{39}$ | 6.69 |
| Pb | 18 | 0.04-0.05 | 0.003 | 6.3 | 37 [a] | 0.42 [a] | 28.9 | 11.3 |
| PbTe | 57 [40] | 0.43 [41] | 0.008 | 22.6 | 40 [a] | 0.26 [40] | 20.3 [42] | 8.16 |
| CdTe | 52 | $0.76^{43}$ | 0.013 | 18.4 | 96 [a] | 0.41 [a] | 5.9 | 5.85 |
| CIGS | 70 [44] | 0.7-1.5 [44] | 0.021-0.010 | 25.9 | 72 [44] | 0.2-0.5 [45] | ~10 [45] | ~5.7 |
| GaAs | $123^{46}$ | $8.4^{46}$ | 0.068 | 46.9 | 107 [a] | 0.31 | 6.4 | 5.32 |
| Si | $169^{46}$ | $12.7^{46}$ | 0.075 | 72.2 | 62 [a] | 0.06-0.28 | 2.6 | 2.33 |

[a] Estimated for a homogeneous, isotropic and linear elastic material, according to[24]: $K = \frac{E}{3(1-2\nu)}$

[b] Estimated for a homogeneous, isotropic and linear elastic material, according to[24]: $G = \frac{E}{2(1+\nu)}$.

[c] Estimated value.

[d] Estimated from speed-of-sound average value of $PbBr_2$ (2282 $\frac{m}{sec}$) [27], using the formula[47]: $v_{s(P)} = \sqrt{\frac{E}{\rho} \cdot \frac{(1-\nu)}{(1+\nu)(1-2\nu)}}$ with an estimated Poisson ratio of 0.33.

[e] The speed of sound- based bulk modulus is 26 GPa. This value was not used as it requires an unreasonably large Poisson ratio (~ 0.4) to match the data.

The bond length and stiffness trends of 'classic' ionic (oxide) perovskite materials are such that the bulk modulus decreases as the ionic radius of the 'A' cuboctahedral group increases[28,29]. For the oxide perovskites this increase of the A radius is usually accompanied by an increase in the unit-cell volume, which is opposite to the case for $APbBr_3$ with 0.198 nm$^3$ for $CsPbBr_3$ and 0.207 nm$^3$ for $CH_3NH_3PbBr_3$. The difference in the volume is also reflected in different distances between the A cation and X anion (see Table 2). A purely electrostatic model would have resulted in a higher elastic modulus for the shorter interatomic bonds, meaning the $CsPbBr_3$. As the results show the opposite, there must be additional forces that keep the hybrid structure bound.

**Table 2**: *Average bond lengths at room temperature vs. the elastic constants of $APbX_3$ bulk materials ($A=CH_3NH_3$, Cs; $X= I$, Br). E and K are Young's and bulk moduli, respectively. The average bond distance refers to structures at room temperature with the mentioned space-groups.*

| Crystal | Average N-X or Cs-X bond distance [Å] | Average Pb-X bond distance [Å] | Space group | E | K |
|---|---|---|---|---|---|
| $CH_3NH_3PbBr_3$ | 4.47 | 2.97 [15] | P-43m | 19.6 | 15.6 |
| $CsPbBr_3$ | 4.15 | 2.96 [16] | Pnma | 15.8 | 15.5 |
| $CH_3NH_3PbI_3$ | 4.50 | 3.16 [36] | I4cm | 14.2 | 12.2 |
| $CsPbI_3$ | 3.90 | 3.23 [36] | Pnma (non-perovskite) | 21 | 19.8 [a] |

[a] Theoretical value [23] – See Table 1.

An interesting similarity to HaPs in sense of their crystal structure (cubic) and surprisingly low thermal conductivity[30,31], can be found in a structural family called *clathrates*. Inorganic clathrate compounds (e.g., $K_8Al_8Si_{36}$, $Sr_8Ga_{16}Ge_{30}$) are characterized by an open framework structure with 'guest' ions (e.g. K, Al and Sr, Ga) trapped in a well-crystallized 'host' framework structure (e.g. Si, Ge). For perovskites, the 'host' can be imagined to be as the Pb-X framework and the guest will be the A ion. In sense of the thermal conductivity – for both cases the main contribution to this effect was related to interactions between the collective 'host' vibrations with the trapped 'guest' atoms/ions [30,32]. As for the elasto-mechanical properties of clathrates, theoretical work showed that the bulk modulus is logarithmically related to the average interatomic distance of the 'host' atoms (See Figure 3 or Ref. [33]). By placing data of the $ABX_3$ (B= Sn or Pb) and $PbX_2$, which are presented in Table 1, and other *oxide* structures found in the literature on a single plot (see Figure 3), we find, for perovskite-structured $APbX_3$, a logarithmic relation between the B-X bond length and the modulus, but with lower elastic modulus than expected.

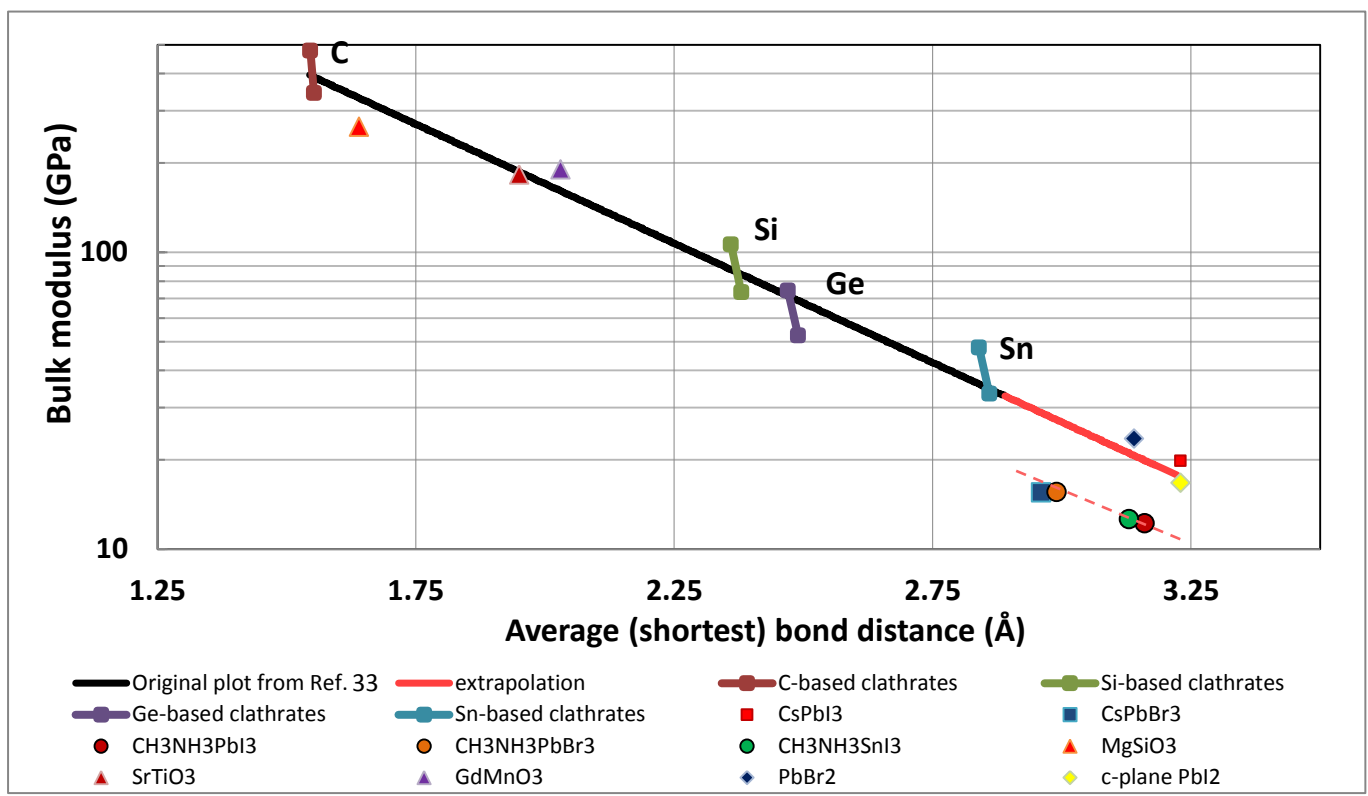

**Figure 3**: Bulk modulus vs. average shortest bond distance. "Average (shortest) bond distance" means: for clathrates - 'host'-'host' average bond distance (e.g., for Si based clathrates this will be the –Si-Si bond); for any $ABX_3$ compound it is the average B-X distance; for $PbX_2$ structures is the average Pb-X distance. The actual numerical values used for the plot are summarized in Table S1. The black line is a linear fit adapted from ref 33.

In order to check the influence of these microscopic considerations on behavior of the bulk material, we looked at the relation between Young's modulus and other physical properties of materials (density, thermal expansion coefficient), with the aid of 2D maps of density or thermal expansion coefficient vs. Young's modulus, from the work of Ashby[24] (see Figs. S1 and S2). On these maps the HaP materials are located between polymers, metals and composite materials.

For a given E, the thermal expansion coefficients of the HaPs are higher than of most metals and composites, smaller than of most polymers and similar to those of rigid polymer foams (and to that of $PbI_2$; see below). It is remarkable that what is formally an ionic material has such a large expansion coefficient, which may reflect the flexibility of the Pb-X bonds. If we compare both the elastic moduli and the linear expansion coefficients of $CH_3NH_3PbX_3$ and $PbX_2$ (c-plane in $PbI_2$ - see Fig. S3), it appears that the Pb-X bond plays a dominant role in the elasticity of these materials.

The finding that HaPs are positioned in the diagrams with relatively soft metals and that the Pb-X corner-sharing octahedral framework is quite (thermally) expandable might indicate possible ductility or flexibility, which could ease formation on heterogeneous surfaces with low energetic penalty and good adaptivity. Another implication of enhanced flexibility is that application of external energy (mechanical, electrostatic, electromagnetic) to such materials might allow facile creation but also ready rearrangement (self-healing) of point defects, which can help atom/ion diffusion in the material. In this context, theoretical calculations gave as the energy barrier to iodide vacancy migration ~0.1 eV [34], up to ~0.3 eV for iodide displacement in a vacancy-free surrounding environment (due to proton migration[35]). Confirmation of suggested processes[6,7,34] such as extensive ion migration and 'self-healing' in these materials awaits further, direct experimental evidence.

If we look at possible benefits of these mechanical properties for manufacturing, these properties can be beneficial for spreading thin films of perovskites on surfaces with different morphologies, using methods such as 'doctor-blading'. However, the ease of plastic deformation (i.e., low hardness), makes these materials less suitable for flexible substrates.

As to any correlation with other lead-based (Pb, PbTe), or photovoltaic materials (CdTe, CIGS, GaAs, Si) – except for the similar elastic modulus for metallic lead, and some similarity to the hardness of the PbTe and CdTe – the overall mechanical properties do not really correlate.

**Summary:**

The elasto-mechanical properties of the examined halide perovskites ($CH_3NH_3PbI_3$, $CH_3NH_3PbBr_3$ and $CsPbBr_3$), with respect to their density or linear thermal expansion coefficient, are between those of soft metals, polymers and composite materials. This combination of properties might be reflected in the (apparent low energies required for) dynamic processes speculated to exist in these materials (e.g. self-healing, ion-migration). Similar to what is the case for 'classic' (oxide) perovskites, the elasto-mechanical properties seem to be dominated by the B-X bond, with the presence of the A group decreasing the elastic coefficients of these large-cage perovskite-structured materials. The lower elastic modulus of $CH_3NH_3PbI_3$ than of $CH_3NH_3PbBr_3$ may reflect differences in the Pb-X bond strengths. Comparison between $CH_3NH_3PbBr_3$ and $CsPbBr_3$ shows that the organic group actually makes the entire structure stiffer (higher elastic modulus). The similarity between the elastic modulus of $PbI_2$ (c plane) with that of $CH_3NH_3PbI_3$ and that of $PbBr_2$ (which has a structure that is quite different from $PbI_2$, which cannot easily be related to a perovskite structure) with $APbBr_3$, suggests that the Pb-X framework dominates the elastic modulus. We do not find an obvious relation between the mechanical and the optoelectronic, especially photovoltaic properties, of the materials.

The conclusions and conjectures presented in this work should mostly serve as guidelines for further investigations, such as growth and mechanics for analogous single crystals (e.g., chloride and formamidinium perovskites).

**Acknowledgements**

DC thanks Leeor Kronik for drawing our attention to this experimental approach. We thank Milko van der Boom for helpful discussions. This work was supported by the Israel Science Foundation, the Israel Ministry of Science and the Israel National Nano-Initiative. DC holds the Sylvia and Rowland Schaefer Chair in Energy Research.

**References:**

1. Kulbak, M., Cahen, D. & Hodes, G. : How Important Is the Organic Part of Lead Halide Perovskite Photovoltaic Cells: Efficient CsPbBr$_3$ Cells. *J. Phys. Chem. Lett.* **6,** 2452–2456 (2015).

2. Green, M. A., Emery, K., Hishikawa, Y., Warta, W. & Dunlop, E. D. : Solar cell efficiency tables (version 44). *Prog. Photovolt. Res. Appl.* **22,** 701–710 (2014).

3. Yin, W.-J., Shi, T. & Yan, Y. : Unusual defect physics in CH$_3$NH$_3$PbI$_3$ perovskite solar cell absorber. *Appl. Phys. Lett.* **104,** 063903 (2014).

4. Shi, D., Adinolfi, V., Comin, R., Yuan, M., Alarousu, E., Buin, A., Chen, Y., Hoogland, S., Rothenberger, A., Katsiev, K., Losovyj, Y., Zhang, X., Dowben, P. A., Mohammed, O. F., Sargent, E. H. & Bakr, O. M. : Low trap-state density and long carrier diffusion in organolead trihalide perovskite single crystals. *Science* **347,** 519–522 (2015).

5. Tress, W., Marinova, N., Moehl, T., Zakeeruddin, S. M., Nazeeruddin, M. K. & Grätzel, M. : Understanding the rate-dependent J–V hysteresis, slow time component, and aging in CH3NH3PbI3 perovskite solar cells: the role of a compensated electric field. *Energy Environ. Sci.* **8,** 995–1004 (2015).

6. Yang, T.-Y., Gregori, G., Pellet, N., Grätzel, M. & Maier, J. : The Significance of Ion Conduction in a Hybrid Organic–Inorganic Lead-Iodide-Based Perovskite Photosensitizer. *Angew. Chem.* **127,** 8016–8021 (2015).

7. Eames, C., Frost, J. M., Barnes, P. R. F., O'Regan, B. C., Walsh, A. & Islam, M. S. : Ionic transport in hybrid lead iodide perovskite solar cells. *Nat. Commun.* **6,** (2015).

8. Xiao, Z., Yuan, Y., Shao, Y., Wang, Q., Dong, Q., Bi, C., Sharma, P., Gruverman, A. & Huang, J. : Giant switchable photovoltaic effect in organometal trihalide perovskite devices. *Nat. Mater.* **14,** 193–198 (2015).

9. Kawamura, Y., Mashiyama, H. & Hasebe, K. : Structural Study on Cubic–Tetragonal Transition of CH$_3$NH$_3$PbI$_3$. *J. Phys. Soc. Jpn.* **71,** 1694–1697 (2002).

10. Mashiyama, H., Kawamura, Y., Magome, E. & Kubota, Y.: Displacive character of the cubic-tetragonal transition in $CH_3NH_3PbX_3$. *J. Korean Phys. Soc.* **42,** S1026–S1029 (2003).

11. Oliver, W. c. & Pharr, G. m. : An improved technique for determining hardness and elastic modulus using load and displacement sensing indentation experiments. *J. Mater. Res.* **7,** 1564–1583 (1992).

12. Dong, Q., Fang, Y., Shao, Y., Mulligan, P., Qiu, J., Cao, L. & Huang, J. : Electron-hole diffusion lengths >175 μm in solution grown $CH_3NH_3PbI_3$ single crystals. *Science* **347,** 967–970 (2015).

13. Dang, Y., Liu, Y., Sun, Y., Yuan, D., Liu, X., Lu, W., Liu, G., Xia, H. & Tao, X. : Bulk crystal growth of hybrid perovskite material $CH_3NH_3PbI_3$. *CrystEngComm* **17,** 665–670 (2014).

14. Tidhar, Y., Edri, E., Weissman, H., Zohar, D., Hodes, G., Cahen, D., Rybtchinski, B. & Kirmayer, S.: Crystallization of Methyl Ammonium Lead Halide Perovskites: Implications for Photovoltaic Applications. *J. Am. Chem. Soc.* **136,** 13249–13256 (2014).

15. Zhao, P., Xu, J., Dong, X., Wang, L., Ren, W., Bian, L. & Chang, A. : Large-Size $CH_3NH_3PbBr_3$ Single Crystal: Growth and In Situ Characterization of the Photophysics Properties. *J. Phys. Chem. Lett.* 2622–2628 (2015)

16. Stoumpos, C. C., Malliakas, C. D., Peters, J. A., Liu, Z., Sebastian, M., Im, J., Chasapis, T. C., Wibowo, A. C., Chung, D. Y., Freeman, A. J., Wessels, B. W. & Kanatzidis, M. G. : Crystal Growth of the Perovskite Semiconductor $CsPbBr_3$: A New Material for High-Energy Radiation Detection. *Cryst. Growth Des.* **13,** 2722–2727 (2013).

17. Li, X. & Bhushan, B. : A review of nanoindentation continuous stiffness measurement technique and its applications. *Mater. Charact.* **48,** 11–36 (2002).

18. Swainson, I. P., Tucker, M. G., Wilson, D. J., Winkler, B. & Milman, V. : Pressure Response of an Organic−Inorganic Perovskite: Methylammonium Lead Bromide. *Chem. Mater.* **19,** 2401–2405 (2007).

19. Hirotsu, S., Suzuki, T. & Sawada, S. : Ultrasonic Velocity around the Successive Phase Transition Points of $CsPbBr_3$. *J. Phys. Soc. Jpn.* **43,** 575 (1977).

20. Rodová, M., Brožek, J., Knížek, K. & Nitsch, K. : Phase transitions in ternary caesium lead bromide. *J. Therm. Anal. Calorim.* **71,** 667–673 (2003).

21. Egger, D. A. & Kronik, L. : Role of Dispersive Interactions in Determining Structural Properties of Organic–Inorganic Halide Perovskites: Insights from First-Principles Calculations. *J. Phys. Chem. Lett.* **5,** 2728–2733 (2014).

22. Feng, J. : Mechanical properties of hybrid organic-inorganic $CH_3NH_3BX_3$ (B = Sn, Pb; X = Br, I) perovskites for solar cell absorbers. *APL Mater.* **2,** 081801 (2014).

23. Murtaza, G. & Ahmad, I. : First principle study of the structural and optoelectronic properties of cubic perovskites $CsPbM_3$ (M=Cl, Br, I). *Phys. B Condens. Matter* **406,** 3222–3229 (2011).

24. Ashby, M. F. *Materials selection in mechanical design*. (Elsevier/Butterworth-Heinemann, 2011).

25. Luo, Y.-R. *Comprehensive Handbook of Chemical Bond Energies*. (CRC Press, 2007).

26. Veiga, W. & Lepienski, C. M. : Nanomechanical properties of lead iodide ($PbI_2$) layered crystals. *Mater. Sci. Eng. A* **335,** 6–13 (2002).

27. Weber, M. J. *CRC Handbook of Laser Science and Technology Supplement 2: Optical Materials*. (CRC Press, 1994).

28. Bhadram, V. S., Swain, D., Dhanya, R., Polentarutti, M., Sundaresan, A. & Narayana, C. : Effect of pressure on octahedral distortions in $RCrO_3$ (R = Lu, Tb, Gd, Eu, Sm): the role of R-ion size and its implications. *Mater. Res. Express* **1,** 026111 (2014).

29. Verma, A. S. & Kumar, A. : Bulk modulus of cubic perovskites. *J. Alloys Compd.* **541,** 210–214 (2012).

30. Pisoni, A., Jaćimović, J., Barišić, O. S., Spina, M., Gaál, R., Forró, L. & Horváth, E. : Ultra-Low Thermal Conductivity in Organic–Inorganic Hybrid Perovskite $CH_3NH_3PbI_3$. *J. Phys. Chem. Lett.* **5,** 2488–2492 (2014).

31. Nakamura, K., Yamada, S. & Ohnuma, T. : Energetic Stability and Thermoelectric Property of Alkali-Metal-Encapsulated Type-I Silicon-Clathrate from First-Principles Calculation. *Mater. Trans.* **54,** 276–285 (2013).

32. Sui, F., He, H., Bobev, S., Zhao, J., Osterloh, F. E. & Kauzlarich, S. M. : Synthesis, Structure, Thermoelectric Properties, and Band Gaps of Alkali Metal Containing Type I Clathrates: $A_8Ga_8Si_{38}$ (A = K, Rb, Cs) and $K_8Al_8Si_{38}$. *Chem. Mater.* **27,** 2812–2820 (2015).

33. Karttunen, A. J., Härkönen, V. J., Linnolahti, M. & Pakkanen, T. A. : Mechanical Properties and Low Elastic Anisotropy of Semiconducting Group 14 Clathrate Frameworks. *J. Phys. Chem. C* **115,** 19925–19930 (2011).

34. Azpiroz, J. M., Mosconi, E., Bisquert, J. & Angelis, F. D. : Defect migration in methylammonium lead iodide and its role in perovskite solar cell operation. *Energy Environ. Sci.* **8,** 2118–2127 (2015).

35. Egger, D. A., Kronik, L. & Rappe, A. M. : Theory of Hydrogen Migration in Organic–Inorganic Halide Perovskites. *Angew. Chem. Int. Ed.* **54,** 1-5 (2015).

36. Stoumpos, C. C., Malliakas, C. D. & Kanatzidis, M. G.: Semiconducting Tin and Lead Iodide Perovskites with Organic Cations: Phase Transitions, High Mobilities, and Near-Infrared Photoluminescent Properties. *Inorg. Chem.* **52,** 9019–9038 (2013).

37. Lee, Y., Mitzi, D. B., Barnes, P. W. & Vogt, T. : Pressure-induced phase transitions and templating effect in three-dimensional organic-inorganic hybrid perovskites. *Phys. Rev. B* **68,** 020103 (2003).

38. Sears, W. M., Klein, M. L. & Morrison, J. A. : Polytypism and the vibrational properties of $PbI_2$. *Phys. Rev. B* **19,** 2305–2313 (1979).

39. Nitsch, K. & Rodová, M. : Thermomechanical Measurements of Lead Halide Single Crystals. *Phys. Status Solidi B* **234,** 701–709 (2002).

40. Ni, J. E., Case, E. D., Khabir, K. N., Stewart, R. C., Wu, C.-I., Hogan, T. P., Timm, E. J., Girard, S. N. & Kanatzidis, M. G. : Room temperature Young's modulus, shear modulus, Poisson's ratio and hardness of PbTe–PbS thermoelectric materials. *Mater. Sci. Eng. B* **170,** 58–66 (2010).

41. Darrow, M. S., White, W. B. & Roy, R. : Micro-indentation hardness variation as a function of composition for polycrystalline solutions in the systems PbS/PbTe, PbSe/PbTe, and PbS/PbSe. *J. Mater. Sci.* **4,** 313–319 (1969).

42. Houston, B., Strakna, R. E. & Belson, H. S. : Elastic Constants, Thermal Expansion, and Debye Temperature of Lead Telluride. *J. Appl. Phys.* **39,** 3913–3916 (1968).

43. Pang, M., Bahr, D. F. & Lynn, K. G. : Effects of Zn addition and thermal annealing on yield phenomena of CdTe and $Cd_{0.96}Zn_{0.04}Te$ single crystals by nanoindentation. *Appl. Phys. Lett.* **82,** 1200–1202 (2003).

44. Luo, S., Lee, J.-H., Liu, C.-W., Shieh, J.-M., Shen, C.-H., Wu, T.-T., Jang, D. & Greer, J. R. : Strength, stiffness, and microstructure of $Cu(In,Ga)Se_2$ thin films deposited via sputtering and co-evaporation. *Appl. Phys. Lett.* **105,** 011907 (2014).

45. Lin, Y.-C., Peng, X.-Y., Wang, L.-C., Lin, Y.-L., Wu, C.-H. & Liang, S.-C. : Residual stress in CIGS thin film solar cells on polyimide: simulation and experiments. *J. Mater. Sci. Mater. Electron.* **25,** 461–465 (2014).

46. Grillo, S. E., Ducarroir, M., Nadal, M., Tournié, E. & Faurie, J.-P. : Nanoindentation of Si, GaP, GaAs and ZnSe single crystals. *J. Phys. Appl. Phys.* **36,** L5 (2003).

47. Fundamentals of Acoustics 4th Ed - L. Kinsler, Et Al., (Wiley, 2000) <https://www.scribd.com/doc/39846878/Fundamentals-of-Acoustics-4th-Ed-L-Kinsler-Et-Al-Wiley-2000-WW-marcado> ; accessed at August 10[th], 2015

# Supplementary Information

# Mechanical Properties of APbX$_3$ (A=Cs or CH$_3$NH$_3$; X=I or Br) Perovskite Single Crystals

Yevgeny Rakita, Sidney R. Cohen* [†], Nir Klein Kedem, Gary Hodes, David Cahen*

Materials and Interfaces Department and [†]Surface Science Laboratory (Chemical Research Support), Weizmann Institute of Science, Rehovot. 76100, Israel

* corresponding authors: sidney.cohen@weizmann.ac.il ; david.cahen@weizmann.ac.il

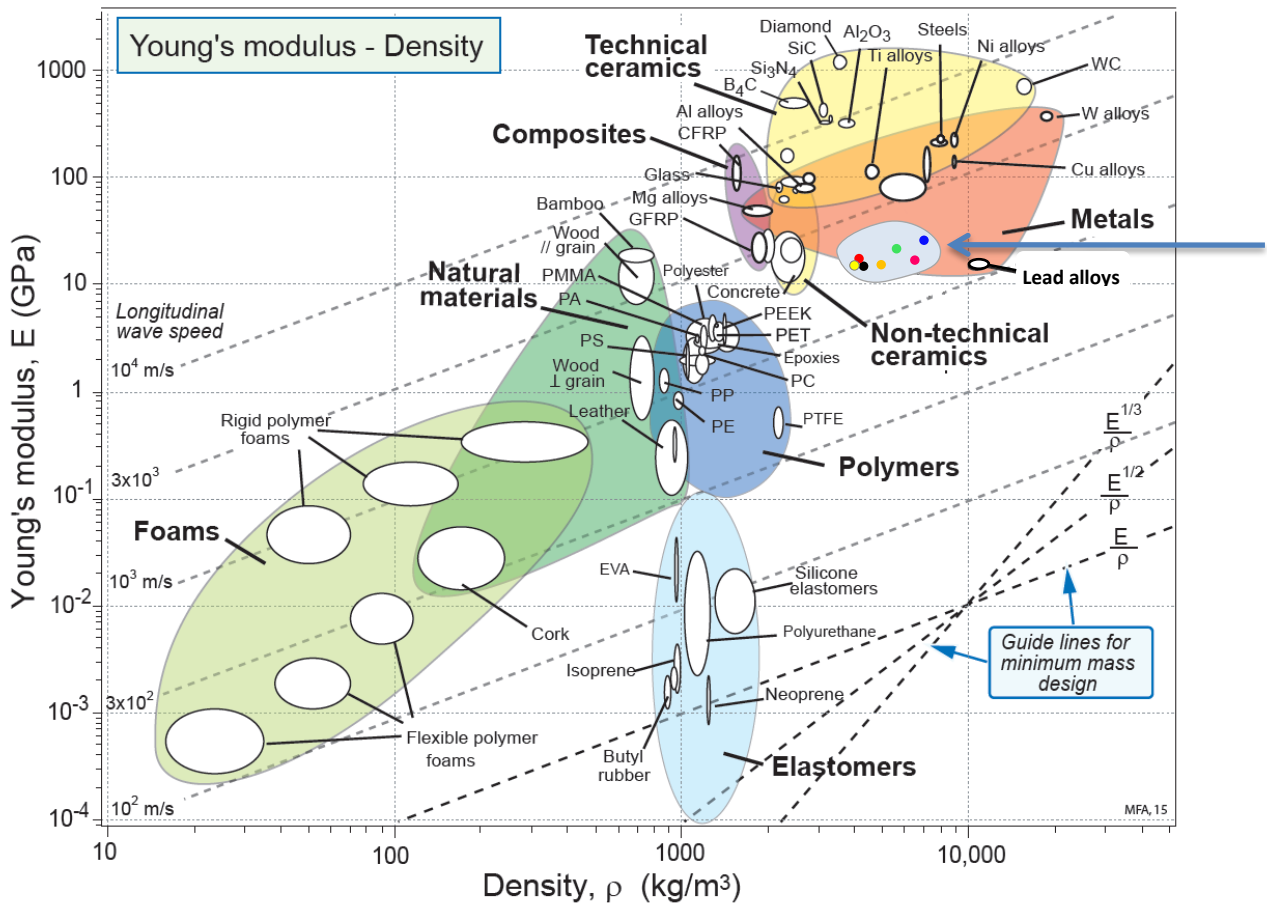

**Figure S2**: Ashby 2D map: density vs. Young's modulus. Experimentally derived values (from this work or other cited works – see Table 1) are added in the area, designated "ABX$_3$" to which the arrow points: **Red** - CH$_3$NH$_3$PbBr; **Black** - CH$_3$NH$_3$PbI$_3$; **Yello** - CH$_3$NH$_3$SnI$_3$; **Orange** - CsPbBr$_3$; **Green** - CsPbI$_3$; **Magenta** – PbI$_2$ ; **Blue** – PbBr$_2$. Reprinted with permission from Ashby, M. F. *Materials Selection in Mechanical Design, Fourth Edition*. (Butterworth-Heinemann, 2010).

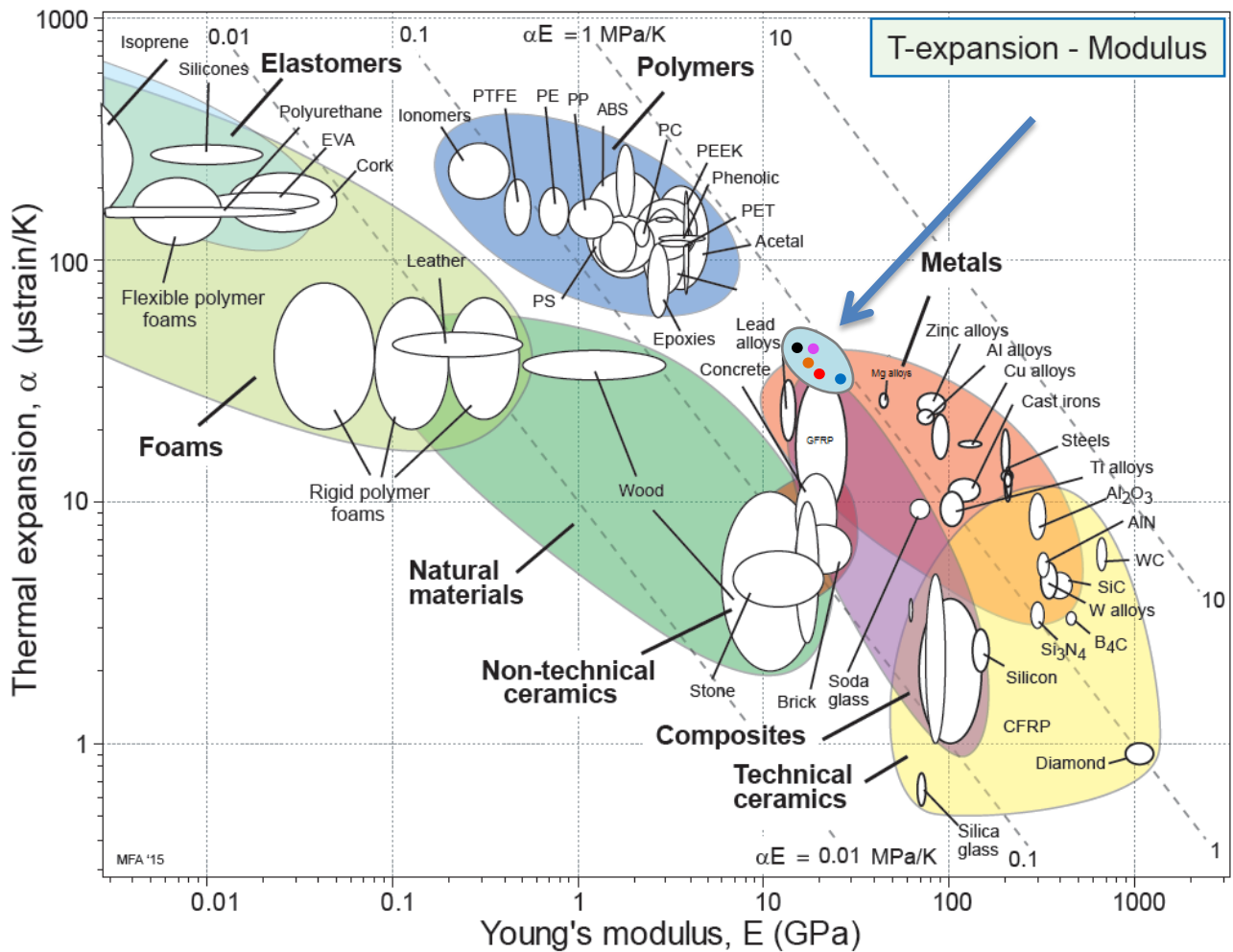

**Figure S3**: Ashby 2D map: linear thermal expansion coefficient vs. Young's modulus. Experimentally derived values (from this work or other cited works – see Table 1) are added in the area, designated "ABX$_3$" to which the arrow points: **Red** - CH$_3$NH$_3$PbBr; **Black** - CH$_3$NH$_3$PbI$_3$; **Orange** - CsPbBr$_3$; **Magenta** – PbI$_2$; **Blue** – PbBr$_2$. Reprinted with permission from Ashby, M. F. *Materials Selection in Mechanical Design, Fourth Edition*. (Butterworth-Heinemann, 2010).

**PbI₂**

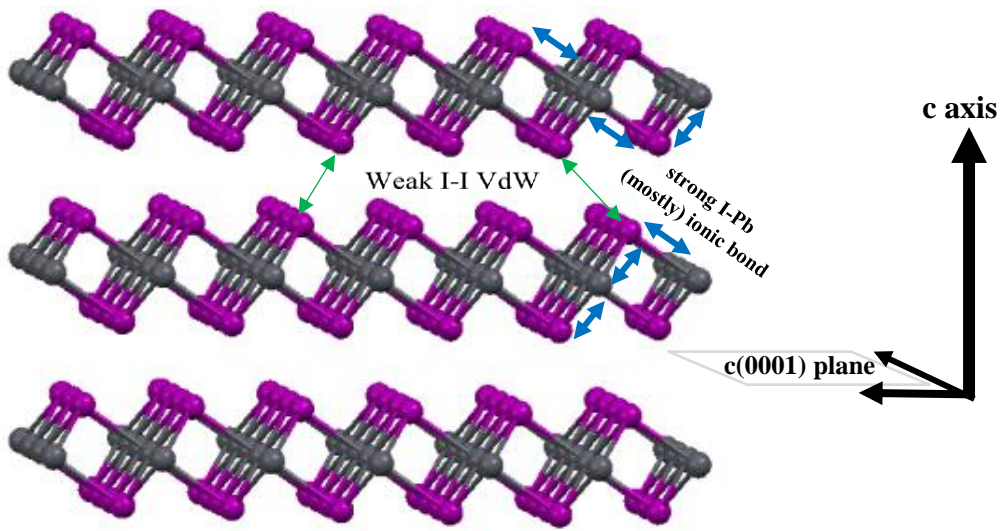

**CH₃NH₃PbI₃**

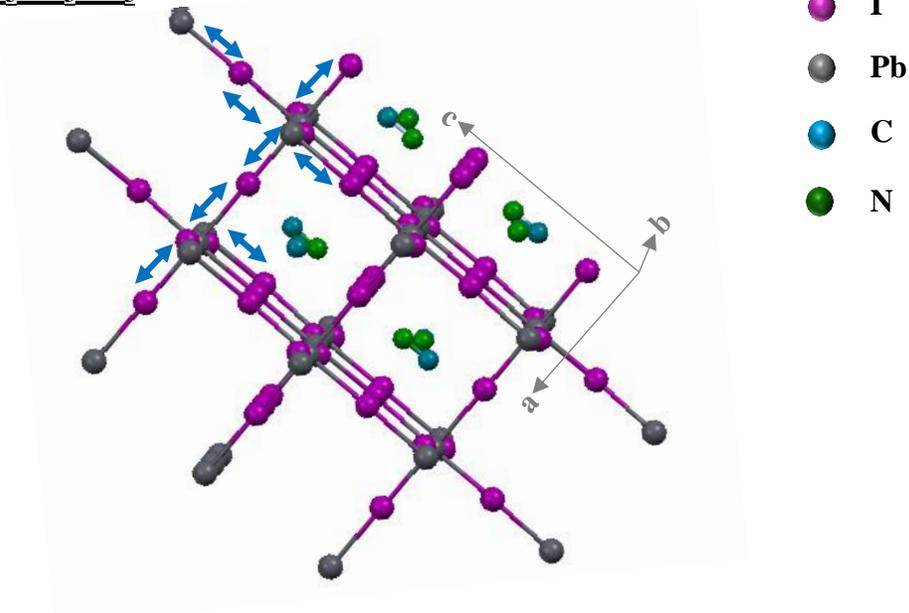

**Figure S4**: Ball-and-stick crystal structure representation of (top) PbI$_2$ and (bottom) CH$_3$NH$_3$PbI$_3$, illustrating the close relation between the Pb-I in the c(0001) plane and in CH$_3$NH$_3$PbI$_3$.

**Table S1**: Values for Figure 3.

| | B-X bond distance (@ RT) [Å] | K [GPa] |
|---|---|---|
| $CH_3NH_3PbI_3$ | 3.16 | 12.2 |
| $CH_3NH_3PbBr_3$ | 2.99 | 15.6 |
| $CH_3NH_3SnI_3$ | 3.13 | 12.6 |
| $CsPbI_3$ | 3.23 | 19.8 |
| $CsPbBr_3$ | 2.96 | 15.5 |
| $MgSiO_3$ | 1.64 | 264 |
| $SrTiO_3$ | 1.95 | 183 |
| $GdMnO_3$ | 2.03 | 190 |
| | | |
| $PbBr_2$ | 3.14 | 23.5 |
| c-plane $PbI_2$ | 3.23 | 16.7 |